\documentclass[aps,prl,twocolumn, groupedaddress]{revtex4-2}
\usepackage{natbib}
\usepackage{parskip}
\usepackage{amsmath}
\usepackage{amssymb}
\usepackage{graphicx}
\usepackage{subcaption}
\usepackage{float}
\usepackage{placeins}
\usepackage[]{caption}
\usepackage{footnote}
\usepackage{geometry}
\title{Stokes drift through corals}
\begin{document}
	\title{Stokes drift through corals}
	\author{Joseph J. Webber}
	\affiliation{Department of Applied Mathematics and Theoretical Physics, University of Cambridge, Wilberforce Road, Cambridge CB3 0WA, UK}
	\altaffiliation{Trinity College, University of Cambridge, Cambridge CB2 1TQ, UK}
	\email[]{jw948@cam.ac.uk}
	\author{Herbert E. Huppert}
	\affiliation{King's College, University of Cambridge, Cambridge CB2 1ST, UK}
	\email{heh1@cam.ac.uk}
	\date{\today}
	\begin{abstract}
		We investigate the all-penetrating drift velocities, due to surface wave motion in an effectively inviscid fluid that overlies a saturated porous bed of finite depth. Previous work in this area either neglects the large-scale flow between layers (Phillips \cite{Phillips1991}) or only considers the drift above the porous layer (Monismith \cite{Monismith2007}). We propose a model where flow is described by a velocity potential above the porous layer, and by Darcy's law in the porous bed, with derived matching conditions at the interface between the two layers. The damping effect of the porous bed requires a complex wavenumber $k$ and both a vertical and horizontal Stokes drift of the fluid, unlike the solely horizontal drift first derived by Stokes \cite{Stokes1847} in a pure fluid layer. Our work provides a physical model for coral reefs in shallow seas, where fluid drift both above and within the reef is vitally important for maintaining a healthy reef ecosystem (Koehl \emph{et al.} \cite{KoehlPowellDobbins1997}, Monismith \cite{Monismith2007}). We compare our model with measurements by Koehl \& Hadfield \cite{KoehlHadfield2004} and also explain the vertical drift effects described in Koehl \emph{et al.} \cite{KoehlStrotherReidenbachKoseffHadfield2007}, who measured the exchange between a coral reef layer and the (relatively shallow) sea above.
	\end{abstract}
	
	\maketitle

	\section{Introduction}
	The importance of wave-driven flows and their role in mass transport within the world's oceans has long been recognised on a macroscopic scale, carrying waste and driftwood \cite{vandenBremerBreivik2017}. One of the key processes which is understood to facilitate this large-scale transport (`drifting') is Stokes drift, a net motion in the direction of wave propagation, first introduced by Stokes \cite{Stokes1847}, arising from the difference between the Lagrangian and Eulerian velocities of fluid particles undergoing oscillatory motion.
	
	The case of Stokes drift arising from surface gravity waves is well-studied, and a detailed exposition of its derivation can be found, for example, in Phillips \cite{Phillips1977}. For a fluid particle with initial position $\boldsymbol{x_0}$ and a Lagrangian velocity field $\boldsymbol{u_L}\left(\boldsymbol{x_0},\,t\right)$, the particle's position at time $t$ is
	\begin{equation}
		\boldsymbol{x}\left(t\right) = \boldsymbol{x_0} + \int^t_0{\boldsymbol{u_L}\left(\boldsymbol{x_0},\,s\right)\,\mathrm{d}s},
	\end{equation}
	and the Eulerian velocity $\boldsymbol{u}\left(\boldsymbol{x},\,t\right)$ is given, upon expansion of a Taylor series, by
	\begin{equation}
		\boldsymbol{u}\left(\boldsymbol{x},\,t\right) = \boldsymbol{u}\left(\boldsymbol{x_0},\,t\right) + \left(\int^t_0{\boldsymbol{u_L}\left(\boldsymbol{x_0},\,s\right)\,\mathrm{d}s}\right) \cdot \boldsymbol{\nabla u} + \dots.
	\end{equation}
	Following the approach of Longuet-Higgins in \cite{LonguetHiggins1969}, this first-order difference term can be time-averaged over a period of oscillation to give the Stokes drift velocity which we will henceforth denote $\boldsymbol{u_S}$. Letting $\left\langle \dots \right\rangle$ denote an average over an oscillation, and dropping the subscript from the Lagrangian velocity, we obtain
	\begin{equation}
		\label{eqn:stokes_drift_velocity}
		\boldsymbol{u_S} = \left\langle \left(\int^t_0{\boldsymbol{u}\left(\boldsymbol{x_0},\,s\right)\,\mathrm{d}s}\right) \cdot \boldsymbol{\nabla u} \right\rangle.
	\end{equation}
	In the case of linear water waves on an ocean surface with mean position $z=0$ and total depth $D$ [described by $z = A\exp{\left\lbrace\mathrm{i}\left(kx - \omega t\right)\right\rbrace}$, where the $x$ coordinate is horizontal], \cite{Phillips1977} shows that the Stokes drift effect is entirely horizontal, with magnitude
	\begin{equation}
		\label{eqn:sd_no_porous_layer}
		u_S = \frac{\omega k A^2 \cosh{\left[2k\left(z+D\right)\right]}}{2\sinh^2{kD}},
	\end{equation}
	which decreases exponentially with depth. However, in many oceanographic contexts, it is not realistic to treat the water as a layer of fixed depth $D$ with an impenetrable boundary at $z=-D$ -- perhaps the simplest example of this is the case of shorelines where the sea overlies a saturated bed of sand, and is known to induce some flow within the sand (as discussed by Phillips \cite{Phillips1991}). One other such potential extension is to a coral reef underlying the ocean. Unlike the dense sand beds discussed by Phillips, flow between the two layers is of great importance as the reef layer is much more permeable on average. A typical coral reef also has a much deeper vertical extent than the porous sand layer typically modelled, before one reaches an effectively solid rock layer below.
	
	The importance of mass transport into and throughout coral reefs is indisputable. Monismith \cite{Monismith2007} states the importance of flow throughout the reef in trapping nutrients and plankton, as well as mass transfer's role in preventing coral bleaching events \cite{NakamuravanWoesik2001}. Further studies have investigated the effects of such flows on larval accumulation \cite{ReidenbachKoseffKoehl2009} and underlined the importance of vertical flows as well as the purely-horizontal effects seen in the absence of a porous layer \cite{KoehlStrotherReidenbachKoseffHadfield2007}.
	
	Fluid-mechanically, however, existing studies of the hydrodynamics of coral reefs tend to consider the porous layer as a boundary condition, exerting drag on the flow above (see, for example, Rosman \& Hench \cite{RosmanHench2011}). Monismith \cite{Monismith2007} uses the model of Longuet-Higgins \& Stewart \cite{LonguetHigginsStewart1962} to treat the effect of propagating waves as a body force on the mean flow, and also cites a number of studies on wave-breaking on the faces of reefs. However, on smaller scales, there has not, to the present authors' knowledge, been any studies on Stokes drift within the coral reef itself, and how the damping effect of this porous layer affects the drift velocities both above and within, aside from the model discussed here and first mentioned in \cite{WebberHuppert2020}.
	
	In this article, we will describe a two-layer model, coupling potential inviscid theory (as first treated in \cite{Stokes1847}) above a viscously-dominated porous layer, where the flow is governed by Darcy's law \cite{Phillips1991}, to derive not only expressions for how incident waves are damped by the presence of a porous layer, but also analytical expressions for the Stokes drift velocities that result. We find that the difference between Eulerian and Lagrangian velocities gives rise not just to a horizontal drift effect, but also to the vertical drifts observed by Koehl \emph{et al.} \cite{KoehlStrotherReidenbachKoseffHadfield2007}, and offer quantitative explanations for this behaviour in terms of the damping of the waves. Finally, we discuss the applicability of the model to complicated real-world coral reefs, and compare the predictions of our model with measurements made in the field.
	
	\section{A two-layer model for a waves over porous media}
\begin{center}
		\begin{figure}[h]
			\centering
			\includegraphics[width=0.9\textwidth]{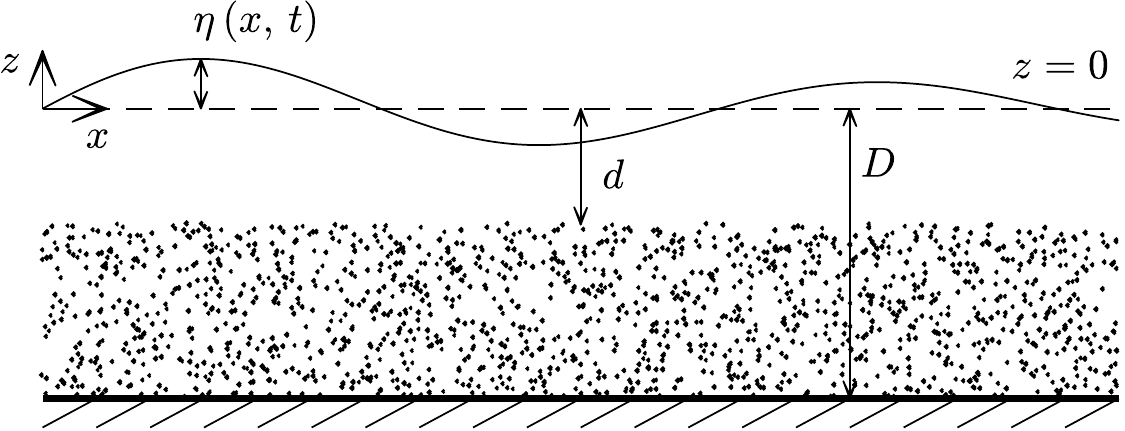}
			\caption{A diagram showing the two-layer configuration under consideration, with a porous medium overlying an impermeable bottom boundary.}
			\label{fig:configuration_diagram}
		\end{figure}
	\end{center}

	We consider a layer of water of total depth $D$ bounded below by an impenetrable floor at $z=-D$. This lower boundary directly underlies a saturated porous medium, of permeability $\Pi\left(z\right)$, which occupies the space between $z=-D$ and $z=-d$. Such a configuration, with surface waves $z=\eta\left(x,\,t\right)$, is shown in figure \ref{fig:configuration_diagram}. In the case $d \to D$ or $\kappa \to \infty$, we recover the classical result derived by Stokes, an example of which is shown in equation (\ref{eqn:sd_no_porous_layer}), which will later serve to provide a check on results.
	\onecolumngrid
	\twocolumngrid
	\subsection{Flow above the porous layer}
	Above the porous layer, we follow \cite{Stokes1847} in letting the fluid velocity be described by a velocity potential $\phi$, such that $\boldsymbol{u} = \boldsymbol{\nabla}\phi$. Incompressibility then imposes that $\nabla^2 \phi = 0$, which is to be solved in $z > -d$ subject to boundary conditions which we will determine.
	
	We start by making the usual assumption that
	\begin{equation}
		\eta{\left(x,\,t\right)} = \mathrm{Re}\left[A e^{\mathrm{i}\left(kx -\omega t\right)}\right],
	\end{equation}
	where the wavenumber $k$ may be complex, to allow for damping effects, which we will later see are key to describing the novel internal vertical drifts driven by horizontal wave motion at the surface. At the free surface $z=\eta\left(x,\,t\right)$, the dynamic boundary condition indicates
	\begin{equation}
		\label{eqn:dynamic_bc_nonlinearised}
		\frac{\partial \phi}{\partial z} =\frac{\partial \eta}{\partial t} + \frac{\partial \phi}{\partial x}\frac{\partial \eta}{\partial x},
	\end{equation}
	which, making the assumptions of linear theory, can be linearised to give $\partial \phi / \partial z = \partial \eta / \partial t$ at $z=0$. This condition implies that $\phi$ must have the same form of $x$- and $t$-dependence as $\eta\left(x,\,t\right)$, such that
	\begin{equation}
		\phi\left(x,\,z,t\right) = \tilde{\phi}\left(z\right)e^{\mathrm{i}\left(kx -\omega t\right)},
	\end{equation}
	where the real part is implicit (henceforth we will always make this assumption). Solving Laplace's equation for $\phi$ with this assumed form allows us to determine $\tilde{\phi}$ up to constants to be determined. Thence,
	\begin{equation}
		\label{eqn:upper_potential}
		\phi\left(x,\,z,t\right) = C_1 \cosh{\left(kz + C_2\right)}e^{\mathrm{i}\left(kx -\omega t\right)},
	\end{equation}
	and the linearised form of equation (\ref{eqn:dynamic_bc_nonlinearised}) shows that
	\begin{equation}
		\label{eqn:const_rel_1}
		k C_1 \sinh{C_2} = -\mathrm{i}\omega A.
	\end{equation}
	Furthermore, we can impose a kinematic boundary condition arising from the Bernoulli equation for unsteady potential flow (see, for example, Batchelor \cite{Batchelor1967}), such that at $z=\eta\left(x,\,t\right)$,
	\begin{equation}
		\rho \frac{\partial \phi}{\partial t} + \frac{\rho}{2}\left|\boldsymbol{\nabla}\phi\right|^2 + \rho g \eta = f\left(t\right),
	\end{equation}
	for $f$ an arbitrary function of time. Without loss of generality, this $f\left(t\right)$ can be set to equal zero, absorbing constants, and so the linearised boundary condition becomes
	\begin{equation}
		\frac{\partial \phi}{\partial t} + g \eta = 0,
	\end{equation}
	providing a second relation between the constants $C_1$ and $C_2$, namely that
	\begin{equation}
		\label{eqn:const_rel_2}
		\mathrm{i} \omega C_1\cosh{C_2} = gA.
	\end{equation}
	
	\subsection{Flow within the porous layer}
	The modelling of flow within porous layers is based on the use of Darcy's law (Phillips \cite{Phillips1991}), relating the volumetric flux to pressure gradients within the fluid. For a porous medium with isotropic permeability $\Pi\left(z\right)$, this flux $\boldsymbol{q}$ is given by
	\begin{equation}
		\boldsymbol{q} = -\frac{\Pi\left(z\right)}{\mu}\boldsymbol{\nabla}p,
	\end{equation}
	where $p$ is the pressure field within the porous medium. In order to facilitate matching with the oscillatory pressure field above the porous medium, we postulate that this pressure field must take the form
	\begin{equation}
		p\left(x,\,z,\,t\right) = \tilde{p}\left(z\right)e^{\mathrm{i}\left(kx -\omega t\right)} + C_3,
	\end{equation}
	for constant $C_3$ and a function $\tilde{p}\left(z\right)$ to be determined. Because we consider incompressible flow, and the matrix of the porous layer is fixed, $\boldsymbol{\nabla}\cdot\boldsymbol{q} = 0$, so $\nabla^2 p = 0$. Further imposing the condition of zero vertical flux across the impermeable boundary at $z=-D$, the pressure field must take the form
	\begin{equation}
	\label{eqn:lower_level_pressure_form}
		p\left(x,\,z,\,t\right) = C_3 + C_4 \cosh{\left[k\left(z+D\right)\right]}e^{\mathrm{i}\left(kx -\omega t\right)}.
	\end{equation}
	Therefore, if $\boldsymbol{q} = \left(u,\,v\right)$,
	\begin{subequations}
		\label{eqn:porous_velocities}
		\begin{align}
			u\left(x,\,z,\,t\right) &= - \frac{\mathrm{i} k \Pi\left(z\right) C_4}{\mu} \cosh{\left[k\left(z+D\right)\right]}e^{\mathrm{i}\left(kx -\omega t\right)}, \\
			v\left(x,\,z,\,t\right) &= -\frac{k \Pi\left(z\right) C_4}{\mu} \sinh{\left[k\left(z+D\right)\right]}e^{\mathrm{i}\left(kx -\omega t\right)}.
		\end{align}
	\end{subequations}

	\subsection{Matching the two layers}
	Having defined our model, we are left with four constants $C_1$, $C_2$, $C_3$ and $C_4$ to be determined. We combine equations (\ref{eqn:const_rel_1}) and (\ref{eqn:const_rel_2}) to find that
	\begin{equation}
		\label{eqn:mc_1}
		\omega^2 = g k \tanh{C_2},
	\end{equation}
	which, as expected, gives the well-known dispersion relation $\omega^2 = gk \tanh{kD}$ (\cite{Lighthill1978}) in the case of no porous layer (fixing $C_2 = kD$ to ensure zero vertical velocity at the lower boundary).
	
	Matching of the inviscid flow above the porous layer and the viscously-dominated Darcy flow within is a non-trivial task. The velocities above the porous layer have a different intrinsic meaning to the spatially-averaged volume fluxes within the porous layer, and it is not immediately clear whether considering the classical matching conditions of velocities and pressures is a valid approach. It is, however, clear that mass conservation requires vertical velocities to match at the interface between the two flow regimes, allowing us to derive the matching condition
	\begin{equation}
		\label{eqn:mc_2}
		- \Pi\left(-d\right) C_4 \sinh{\left[k\left(D-d\right)\right]} = \mu C_1 \sinh{\left(C_2 - kd\right)}.
	\end{equation}
	
	It is clear that we should not match tangential velocities at the interface, however. Beavers \& Joseph \cite{BeaversJoseph1967} remark that there is a `slip' discontinuity between layers, and the forces exerted by the reef on the flow could complicate matters, for example by imposing shear stresses (as is the case in many existing models such as Rosman \& Hench \cite{RosmanHench2011}). The model in \cite{BeaversJoseph1967} also requires continuity of pressure at the interface, and the linearisied Navier-Stokes equations suggest, in $z\ge-d$,
	\begin{equation}
		\rho \frac{\partial \boldsymbol{u}}{\partial t} = -\boldsymbol{\nabla}p - \rho g \boldsymbol{\hat{z}} \quad \text{so} \quad \mathrm{i}\rho \omega \boldsymbol{\nabla} \phi = \boldsymbol{\nabla} p + \rho g \boldsymbol{\hat{z}}.
	\end{equation}
	Then, if we take the (constant) pressure above the surface waves to be zero,
	\begin{equation}
		p = \rho g d + \mathrm{i}\rho \omega C_1 \cosh{\left(C_2 - kd\right)} e^{\mathrm{i}\left(kx -\omega t\right)} \quad \text{at $z=-d$.}
	\end{equation}
	Hence, matching this with the expression in (\ref{eqn:lower_level_pressure_form}), it is seen that $C_3 = \rho g d$ and
	\begin{equation}
		\label{eqn:mc_3}
		C_4 \cosh{\left[k\left(D-d\right)\right]} = \mathrm{i}\rho \omega C_1 \cosh{\left(C_2 - kd\right)}.
	\end{equation}
	
	\subsubsection{Deriving the dispersion relation}
	This condition now means that the solution can be fully-determined. Starting from (\ref{eqn:mc_1}), which determines $C_2$, we can also determine $C_1$ from (\ref{eqn:const_rel_2}), namely that
	\begin{equation}
		C_1 = -\frac{\mathrm{i}A}{\omega k}\sqrt{\left(gk\right)^2 - \omega^4} \quad \text{and} \quad C_2 = \mathrm{arctanh}\left(\frac{\omega^2}{gk}\right).
	\end{equation}
	It is then straightforward to determine the value of $C_4$ using either (\ref{eqn:mc_2}) or (\ref{eqn:mc_3}). This gives
	\begin{equation}
		C_4 = -\frac{\mathrm{i} \mu A \sinh{\left[\mathrm{arctanh}\left(\frac{\omega^2}{gk}\right) - kd\right]}}{\omega k \Pi\left(-d\right) \sinh{\left[k\left(D-d\right)\right]}} \sqrt{\left(gk\right)^2 - \omega^4}.
	\end{equation}
	Perhaps of more interest, however, (\ref{eqn:mc_2}) and (\ref{eqn:mc_3}) can be divided to give
	\begin{equation}
		\label{eqn:disp_rel_raw}
		\mathrm{i} \rho \omega \Pi_I\tanh{\left[k\left(D-d\right)\right]} = -\mu \tanh{\left[\mathrm{arctanh}\left(\frac{\omega^2}{gk}\right) - kd\right]},
	\end{equation}
	where $\Pi_I = \Pi\left(-d\right)$ is the interfacial permeability. This is a dispersion relation linking the frequency $\omega$ of surface waves to their complex wavenumber $k$. It is important to note from the outset that only the value of permeability at the porous layer interface, $\Pi_I$, is of any importance in this relation; this is an artefact of the fact that there is no stress matching condition and instead all matching conditions depend only on values at the boundary.
	
	\subsection{Special cases of the dispersion relation}
	Define the dimensionless constant $J = \rho \omega \Pi\left(-d\right) / \mu $ such that the dispersion relation of equation (\ref{eqn:disp_rel_raw}) can be written
	\begin{equation}
		\label{eqn:disp_rel}
		J \tanh{\left[k\left(D-d\right)\right]} = \mathrm{i} \tanh{\left[\mathrm{arctanh}\left(\frac{\omega^2}{gk}\right) - kd\right]}.
	\end{equation}
	Owing to the nature of this dispersion relation, it will usually need to be solved numerically for $k$ given $\omega$ and the parameters of the model. To investigate the properties of the relation, define
	\begin{align}
		\label{eqn:f_definition}
		f\left(k,\,\omega;\,J,\,D,\,d,\,g\right) =&\, J \tanh{\left[k\left(D-d\right)\right]}\notag\\ &- \mathrm{i} \tanh{\left[\mathrm{arctanh}\left(\frac{\omega^2}{gk}\right) - kd\right]},
	\end{align}
	such that, for fixed $\omega$, zeros of $f$ correspond to suitable values of the wavenumber $k$. Making the assumption that waves propagate in the positive $x$-direction and are therefore damped in this direction (i.e. their amplitude decreases as $x$ increases), we will only seek (the physically-relevant) solutions with $\mathrm{Re}\left(k\right) > 0$ and $\mathrm{Im}\left(k\right) > 0$.
	
\begin{figure}
		\centering
		\begin{subfigure}{0.85\linewidth}
			\centering
			\includegraphics[width=\linewidth]{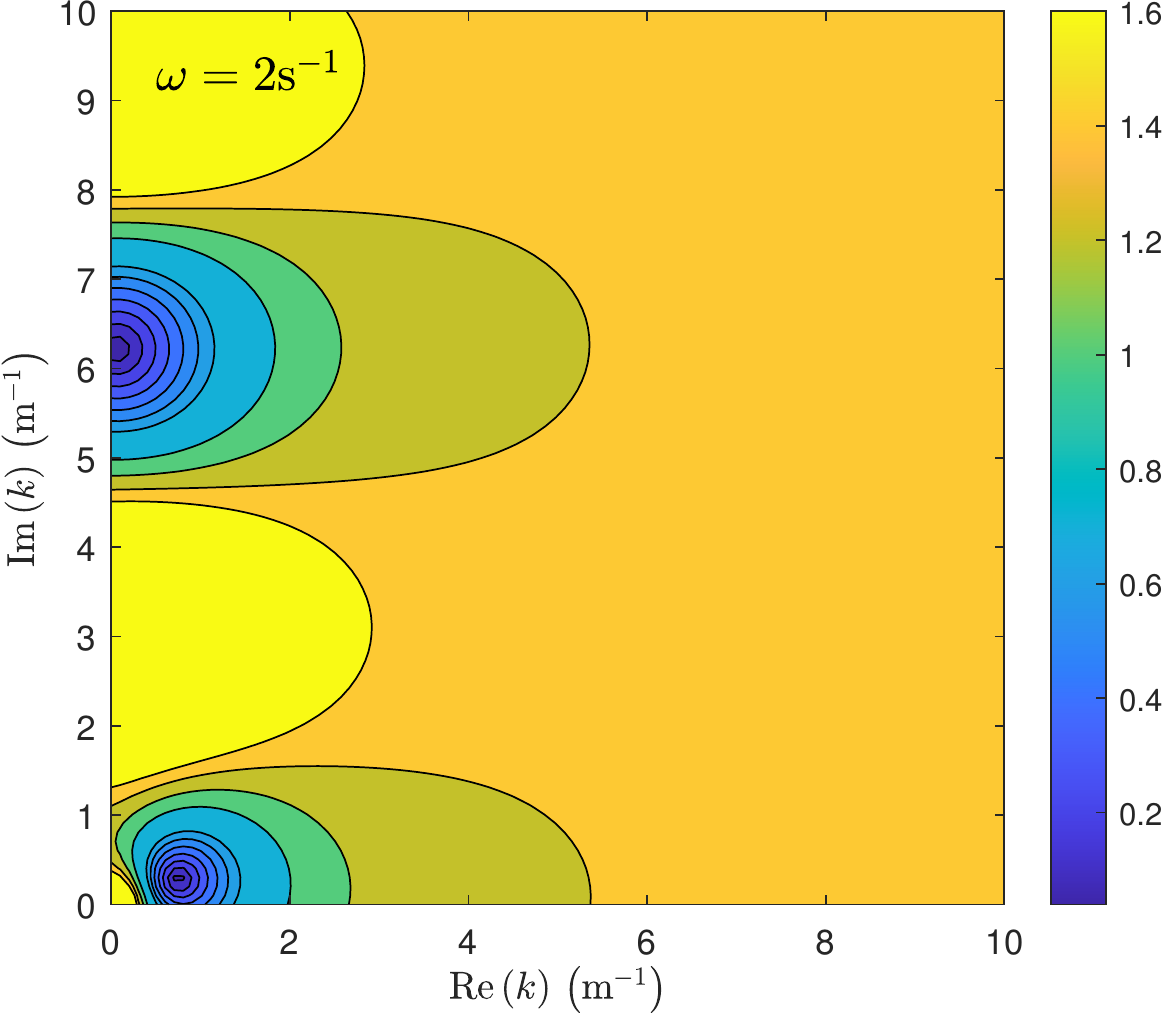}
		\end{subfigure}
		\begin{subfigure}{0.85\linewidth}
			\centering
			\includegraphics[width=\linewidth]{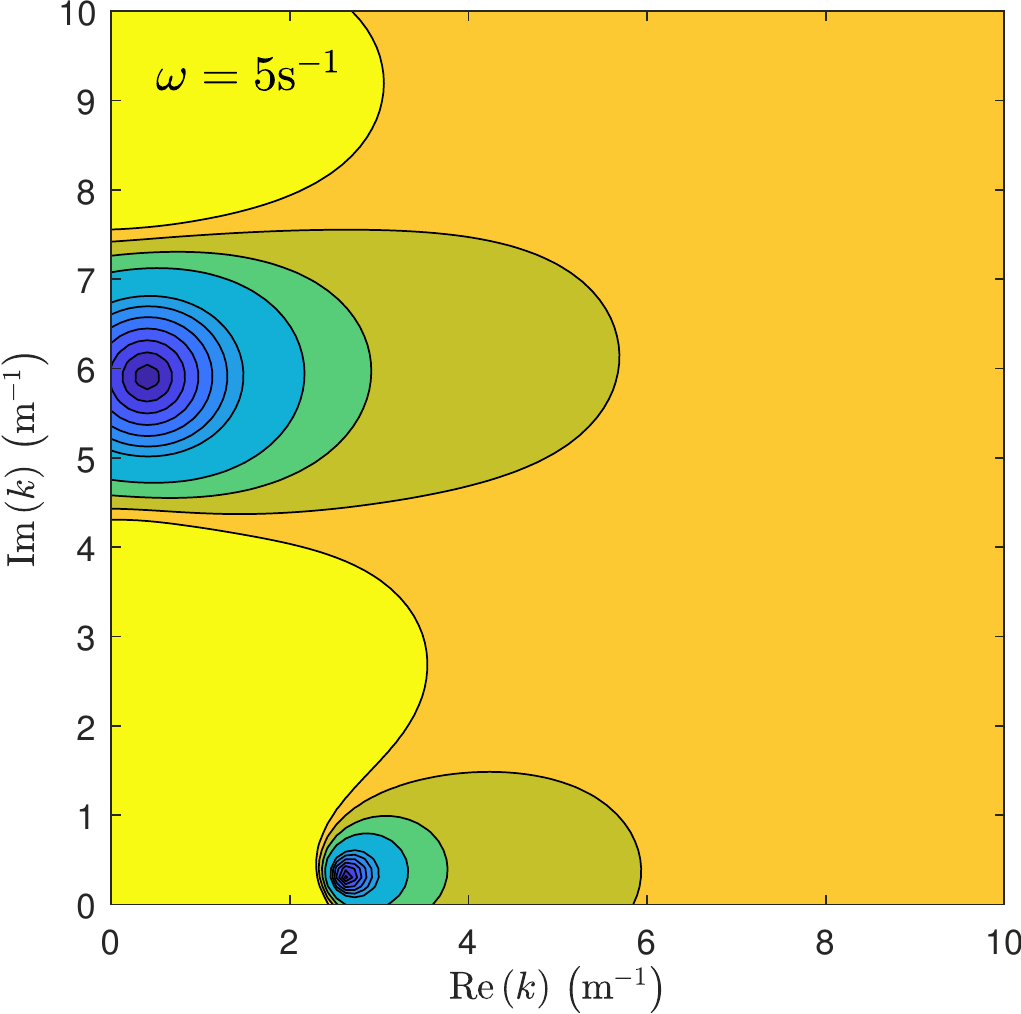}
		\end{subfigure}
		\begin{subfigure}{0.85\linewidth}
			\centering
			\includegraphics[width=\linewidth]{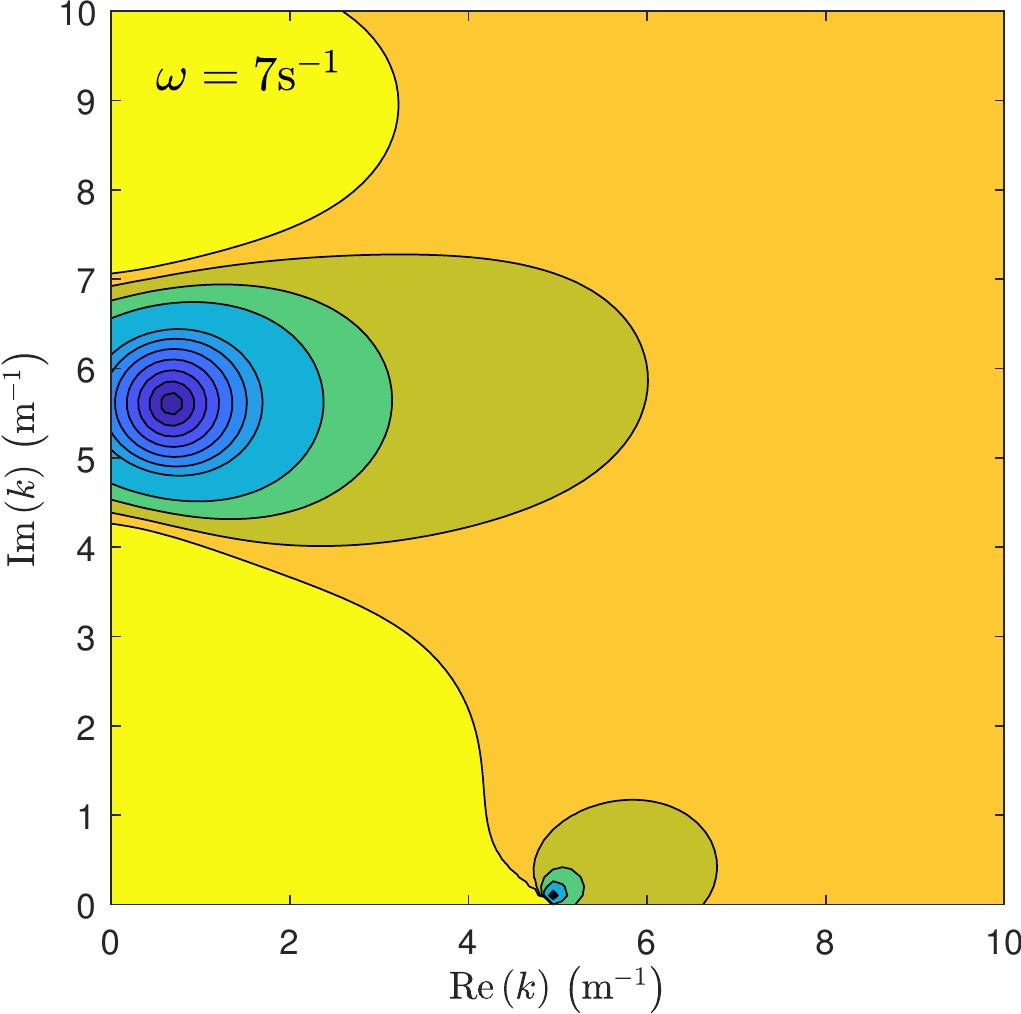}
		\end{subfigure}
		\caption{Contour plots of the function $\left|f\left(k,\,\omega;\,1,\,1,\,1/2,\,10\right)\right|$, with $f$ as defined by (\ref{eqn:f_definition}), showing two solutions for $k$ in each case. Blue represents small magnitudes.}
		\label{fig:contour_plots}
	\end{figure}

	An initial investigation shows that it is possible to find multiple solutions for $k$ that satisfy this assumption, even with all of the other parameters fixed. Figure \ref{fig:contour_plots} shows that there exists one \emph{propagating} solution with a small imaginary part and larger real part, and one \emph{almost evanescent} solution with a very small real part and larger imaginary part. For our purposes here, we will consider the \emph{propagating} solutions because we would expect the \emph{almost evanescent} solutions to decay quickly as they are damped over the reef.

	\subsubsection{Limit of no porous layer}
	\label{sec:disp_rel_no_porous}
	As mentioned above, one would expect that the classical dispersion relation $\omega^2 = gk \tanh{kd}$ \cite{Lighthill1978} would be recovered in the limit of no porous layer. Indeed, taking either $d \to D$ or $\Pi\left(-d\right) = 0$ (i.e. setting the porous layer interface to have zero permeability) sets the left-hand side of equation (\ref{eqn:disp_rel}) to zero, such that
	\begin{equation}
		\tanh{\left[\mathrm{arctanh}\left(\frac{\omega^2}{gk}\right) - kd\right]} \Rightarrow \omega^2 = gk \tanh{kd}.
	\end{equation}
	
	\subsubsection{Limit of small frequencies}
	\label{sec:small_freq}
	In the case of low-frequency waves, we would expect $k$ to be small, giving longer wavelengths. Using the addition formula for $\tanh$,
	\begin{equation}
		\label{eqn:addition_formula}
		J \tanh{\left[k\left(D-d\right)\right]} = \mathrm{i} \frac{\omega^2 / gk - \tanh{\left(kd\right)}}{1 - \omega^2 \tanh{\left(kd\right)}/gk},
	\end{equation}
	and $\tanh{x} \approx x$ for small $x$, so
	\begin{equation}
		Jk\left(D-d\right) \approx \mathrm{i}\frac{\omega^2 - gdk^2}{gk - dk\omega^2},
	\end{equation}
	simplifying to
	\begin{equation}
		\label{eqn:small_freq_lim}
		k \approx \left(\frac{\mathrm{i}\omega^2}{J\left(D-d\right)(g - d\omega^2) + \mathrm{i}gd}\right)^{1/2}.
	\end{equation}
	
	\subsubsection{Limit of high frequencies}
	\label{sec:high_freq}
	Conversely, we would expect $k \to \infty$ as $\omega \to \infty$ and therefore, again starting from equation (\ref{eqn:addition_formula}),
	\begin{equation}
		\label{eqn:high_freq_lim}
		J \approx \mathrm{i}\left(\omega^2 - gk\right)/\left(gk - \omega^2\right),
	\end{equation}
	but because $J$ is real, $\omega^2 = gk$, or, alternatively, $k = \omega^2 / g$. Note here that $k$ is real in this limit, so the asymptotic expression for $\mathrm{Im}\left(k\right)$ is simply zero to leading order. Considering the next order, as detailed in the appendix, gives the asymptotic expansion
	\begin{equation}
		\mathrm{Im}\left(k\right) \approx \frac{2 J \omega^2}{g\left(1+J^2\right)}\left(E_1 + E_2\right),
	\end{equation}
	where $E_1 = e^{-2 \omega^2 d / g}$ and $E_2 = e^{-2\omega^2 \left(D-d\right) / g}$.
	\begin{figure}
		\centering
		\vspace{1em} %Necessary for layout
		\begin{subfigure}{\linewidth}
			\centering
			\includegraphics[width=\linewidth]{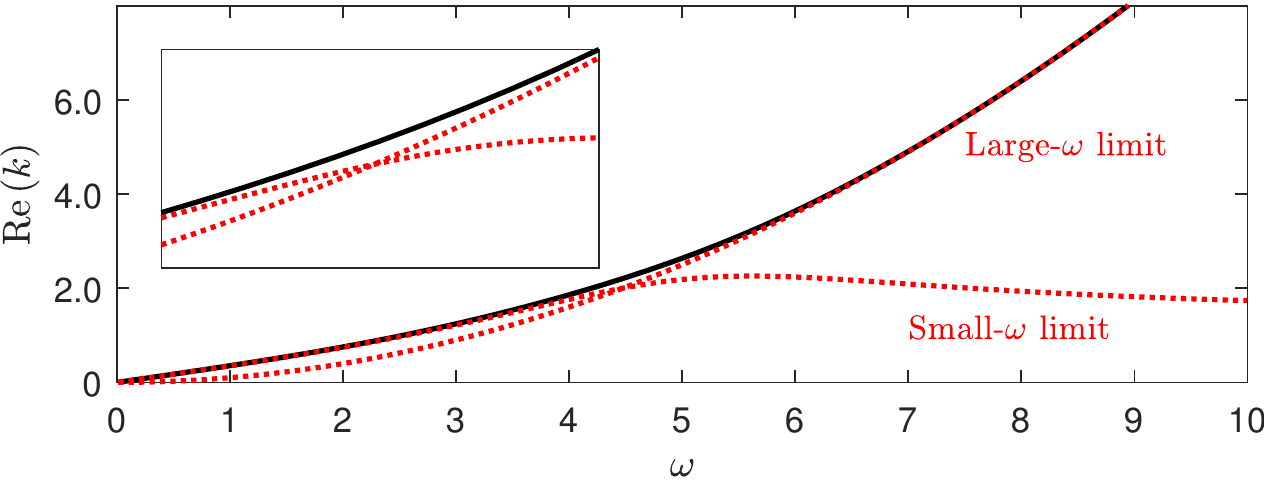}
		\end{subfigure}
		\par\bigskip
		\par\bigskip
		\begin{subfigure}{\linewidth}
			\centering
			\includegraphics[width=\linewidth]{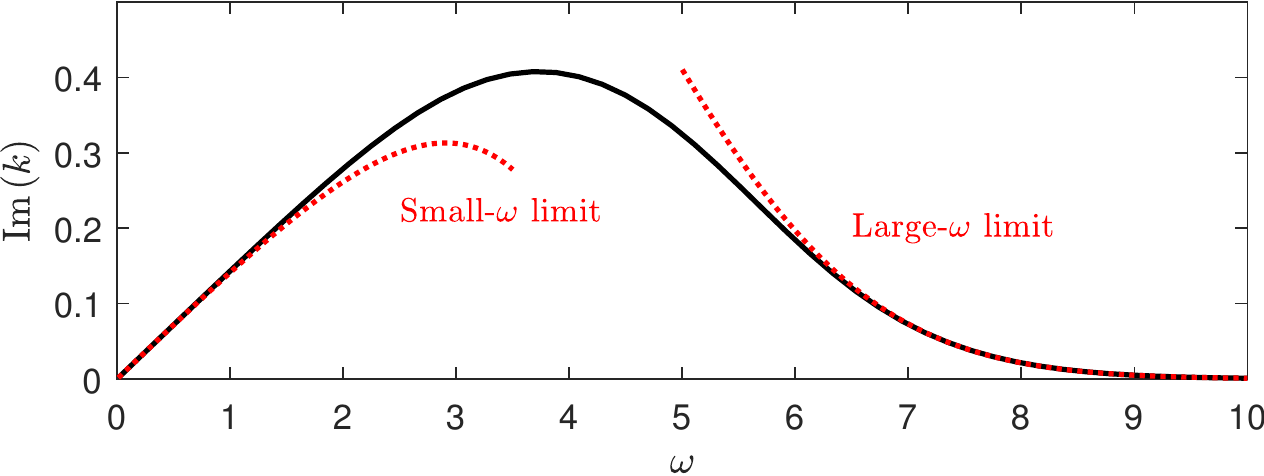}
		\end{subfigure}
		\caption{Plots of the real and imaginary parts of $k$ against $\omega$ in the case $S=1$, $D=1\,\mathrm{m}$, $d=0.5\,\mathrm{m}$ and $g=10\,\mathrm{ms}^{-2}$. The numerical approximations of (\ref{eqn:small_freq_lim}) and (\ref{eqn:high_freq_lim}) are plotted in red, showing close comparison. The inset on the first figure shows the exchange in validity between the large-$\omega$ and small=$\omega$ limits.}
		\label{fig:dispersion_plots}
	\end{figure}

	\section{Stokes drift velocities}
	Following the approach detailed in \cite{Phillips1977}, given the expressions for velocities both above and within the porous layer, it is possible to calculate the Stokes drift velocities in these layers. Provided that particles do not cross between these two layers, a complication which will be later discussed in more detail, this gives an indication of the wave-driven fluid velocities.
	
	\onecolumngrid
	\subsection{Above the porous layer}
	Given the expression of (\ref{eqn:upper_potential}) for the velocity potential, it can be seen that
	\begin{equation}
		\int^t_0{\boldsymbol{u}\,\mathrm{d}s} = \mathrm{Re}\Bigg[\frac{\mathrm{i} k C_1 \left\lbrace e^{\mathrm{i}\left(kx -\omega t\right)} - e^{\mathrm{i}kx}\right\rbrace}{\omega}\Big\lbrace\mathrm{i}\cosh{\left(kz+C_2\right)}\boldsymbol{\hat{x}} + \sinh{\left(kz+C_2\right)}\boldsymbol{\hat{z}}\Big\rbrace\Bigg],
	\end{equation}
	taking real parts as discussed above. Also, again taking real parts,
		\begin{subequations}
			\begin{align}
				\frac{\partial\boldsymbol{u}}{\partial x} &= \mathrm{Re}\left[k^2 C_1 e^{\mathrm{i}\left(kx -\omega t\right)} \left\lbrace-\cosh{\left(kz+C_2\right)}\boldsymbol{\hat{x}} + \mathrm{i}\sinh{\left(kz+C_2\right)}\boldsymbol{\hat{z}}\right\rbrace\right] \; \text{and} \\\frac{\partial\boldsymbol{u}}{\partial z} &= \mathrm{Re}\left[k^2 C_1 e^{\mathrm{i}\left(kx -\omega t\right)} \left\lbrace\mathrm{i}\sinh{\left(kz+C_2\right)}\boldsymbol{\hat{x}} + \cosh{\left(kz+C_2\right)}\boldsymbol{\hat{z}}\right\rbrace\right].
			\end{align}
		\end{subequations}
	Using the result that $\left\langle \mathrm{Re}\left[a e^{\mathrm{i}\omega t}\right] \mathrm{Re}\left[b e^{\mathrm{i}\omega t} + c\right] \right\rangle = \frac{1}{2} \mathrm{Re}\left[a\bar{b}\right]$, where $a$, $b$ and $c$ are time-independent complex numbers,  and equation (\ref{eqn:stokes_drift_velocity}), we can find the Stokes drift velocity above the porous layer
	\begin{equation}
		\boldsymbol{u_S} = \frac{\left|C_1\right|^2 \left|k\right|^2 e^{-2 \mathrm{Im}\left[k\right]x}}{2 \omega} \mathrm{Re} \Bigg[k\left\lbrace\left|\cosh\right|^2 + \left|\sinh\right|^2\right\rbrace\boldsymbol{\hat{x}} - \mathrm{i}k\left\lbrace\cosh\overline{\sinh} +\overline{\cosh}\sinh\right\rbrace\boldsymbol{\hat{z}}\Bigg],
	\end{equation}
	where the arguments of $\cosh$ and $\sinh$, both $kz + C_2$, are omitted for brevity. Then, we can use the identities
	\begin{align}
		\left|\cosh{x}\right|^2 + \left|\sinh{x}\right|^2 = \cosh{\left(2 \mathrm{Re}\left[x\right]\right)} \; \text{and} \notag \\
		\quad \cosh{x}\,\overline{\sinh{x}} +\overline{\cosh{x}}\sinh{x} = \sinh{\left(2 \mathrm{Re}\left[x\right]\right)}
	\end{align}
	to see that
	\begin{equation}
		\label{eqn:sd_upper}
		\boldsymbol{u_S} = \frac{\left|C_1\right|^2 \left|k\right|^2 e^{-2 \mathrm{Im}\left[k\right]x}}{2 \omega} \Big\lbrace \mathrm{Re}\left[k\right] \cosh{\left(2 \mathrm{Re}\left[kz+C_2\right]\right)}\boldsymbol{\hat{x}} + \mathrm{Im}\left[k\right] \sinh{\left(2 \mathrm{Re}\left[kz+C_2\right]\right)}\boldsymbol{\hat{z}}\Big\rbrace.
	\end{equation}

	\subsection{Within the porous layer}
	Analogously within the porous medium, recall that
	\begin{equation}
		\boldsymbol{q} = -\mathrm{Re}\Bigg[\frac{k \Pi\left(z\right) C_4 e^{\mathrm{i}\left(kx-\omega t\right)}}{\mu}\Big\lbrace \mathrm{i}\cosh{\left[k\left(z+D\right)\right]}\boldsymbol{\hat{x}} + \sinh{\left[k\left(z+D\right)\right]}\boldsymbol{\hat{z}} \Big\rbrace\Bigg]
	\end{equation}
	is an analogue for velocity $\boldsymbol{u}$, so
	\begin{subequations}
		\begin{align}
			\int^t_0{\boldsymbol{q}\,\mathrm{d}s} = &-\mathrm{Re}\left[\frac{\mathrm{i} k \Pi\left(z\right) C_4 \left\lbrace e^{\mathrm{i}\left(kx -\omega t\right)} - e^{\mathrm{i}kx}\right\rbrace}{\mu \omega}\left\lbrace\mathrm{i}\cosh{\left[k\left(z+D\right)\right]}\boldsymbol{\hat{x}} + \sinh{\left[k\left(z+D\right)\right]}\boldsymbol{\hat{z}}\right\rbrace\right], \\
			\frac{\partial \boldsymbol{q}}{\partial x} = &-\mathrm{Re}\left[\frac{k^2 \Pi\left(z\right) C_4 e^{\mathrm{i}\left(kx-\omega t\right)}}{\mu}\left\lbrace -\cosh{\left[k\left(z+D\right)\right]}\boldsymbol{\hat{x}} + \mathrm{i}\sinh{\left[k\left(z+D\right)\right]}\boldsymbol{\hat{z}} \right\rbrace\right] \; \text{and} \\
			\frac{\partial \boldsymbol{q}}{\partial z} = &-\mathrm{Re}\left[\frac{k^2 \Pi\left(z\right) C_4 e^{\mathrm{i}\left(kx-\omega t\right)}}{\mu}\left\lbrace \mathrm{i}\sinh{\left[k\left(z+D\right)\right]}\boldsymbol{\hat{x}} + \cosh{\left[k\left(z+D\right)\right]}\boldsymbol{\hat{z}} \right\rbrace\right] \notag \\
			&-\mathrm{Re}\left[\frac{k \Pi'\left(z\right) C_4 e^{\mathrm{i}\left(kx-\omega t\right)}}{\mu}\left\lbrace \mathrm{i}\cosh{\left[k\left(z+D\right)\right]}\boldsymbol{\hat{x}} + \sinh{\left[k\left(z+D\right)\right]}\boldsymbol{\hat{z}} \right\rbrace\right].
		\end{align}
	\end{subequations}
	Then,
	\begin{align}
		\boldsymbol{u_S} = \frac{\left|C_4\right|^2 \left|k\right|^2 \Pi^2\left(z\right) e^{-2 \mathrm{Im}\left[k\right]x}}{2 \mu^2 \omega} \Big\lbrace \mathrm{Re}\left[k\right] \cosh{\left(2 \mathrm{Re}\left[k\left(z+D\right)\right]\right)}\boldsymbol{\hat{x}} + \mathrm{Im}\left[k\right] \sinh{\left(2 \mathrm{Re}\left[k\left(z+D\right)\right]\right)}\boldsymbol{\hat{z}}\Big\rbrace \notag \\
		+\frac{\left|C_4\right|^2 \left|k\right|^2 \Pi\left(z\right) \Pi'\left(z\right) e^{-2 \mathrm{Im}\left[k\right]x}}{2 \mu^2 \omega} \mathrm{Re}\left[\overline{\cosh{\left[k\left(z+D\right)\right]}}\sinh{\left[k\left(z+D\right)\right]}\boldsymbol{\hat{x}} + \mathrm{i}\left|\sinh{\left[k\left(z+D\right)\right]}\right|^2\boldsymbol{\hat{z}}\right].
	\end{align}
	Thus, we can write the two components $\boldsymbol{u_S} = \left(u_S, v_S\right)$ in the form
	\begin{subequations}
		\label{eqn:sd_lower}
		\begin{align}
		u_S &= \frac{\left|C_4\right|^2 \left|k\right|^2 \Pi\left(z\right) e^{-2 \mathrm{Im}\left[k\right]x}}{2 \mu^2 \omega} \Big\lbrace \Pi\left(z\right) \mathrm{Re}\left[k\right] \cosh{\left(2 \mathrm{Re}\left[k\left(z+D\right)\right]\right)} + \frac{\Pi'\left(z\right)}{2} \sinh{\left(2\mathrm{Re}\left[k\left(z+D\right)\right]\right)} \Big\rbrace \\
		\intertext{and}
		v_S &= \frac{\left|C_4\right|^2 \left|k\right|^2 \Pi^2\left(z\right) e^{-2 \mathrm{Im}\left[k\right]x}}{2 \mu^2 \omega} \mathrm{Im}\left[k\right] \sinh{\left(2 \mathrm{Re}\left[k\left(z+D\right)\right]\right)}.
		\end{align}
	\end{subequations}
\vspace{1em}
	\twocolumngrid
	
	\subsection{Interpretation of the results}
\begin{figure}
		\centering
		\includegraphics[width=\linewidth]{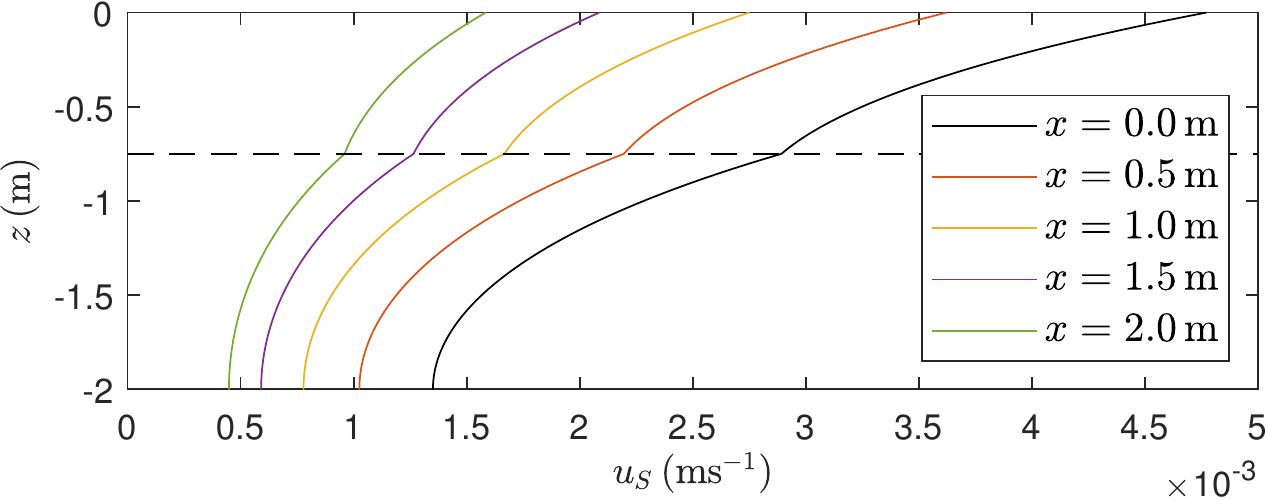}
		\caption{Horizontal Stokes drift velocity profiles for the model with parameters detailed in table \ref{tab:params} measured at different horizontal positions. The damping action of the porous layer results in a reduction in wave amplitude, in turn reducing the magnitude of drift velocities.}
		\label{fig:damping}
	\end{figure}

	Many parallels are seen between the expressions of equations (\ref{eqn:sd_upper}) and (\ref{eqn:sd_lower}) and the classical expressions first derived by \cite{Stokes1847}. In both cases, the magnitudes of the drift velocities scale like the square of the amplitude of fluid motion, with the damping effect of the porous layer manifested in the $e^{-2\mathrm{Im}\left[k\right]x}$ term. This damping can be seen very clearly in figure \ref{fig:damping}, which considers waves with amplitude $0.05\,\mathrm{m}$ at position $x=0$.
	
	It is also of interest to consider the case where $d\to D$ and there is no porous layer. As was seen in the case where there is no porous layer, the dispersion relation becomes $\omega^2 = gk \tanh{kd}$ and thus $C_2 = kd$. This means that
	\begin{equation}
		C_1 = -\mathrm{i}gA/\left(\omega \cosh{kd}\right) \; \text{and} \; C_4 = \rho gA/\left(\cosh{kd}\right).
	\end{equation}
	Then, above the porous layer, equation (\ref{eqn:sd_upper}) becomes
	\begin{equation}
		\boldsymbol{u_S} = \frac{g^2 A^2 k^3}{2 \omega^3 \cosh^2{kd}} \cosh{\left[2k\left(z+d\right)\right]},
	\end{equation}
	which simplifies precisely to the form of (\ref{eqn:sd_no_porous_layer}) in this limit, as would be expected.
	
	\section{Stokes drift in coral reefs}
	It becomes useful at this stage to introduce representative dimensional paramters -- dependence on properties of the porous layer such as the permeability and its geometry cannot be encapsulated in a simple dimensionless parameter like $J$ above. As discussed in the introduction, an understanding of the hydrodynamics of coral reefs is of great importance, and there is a wealth of data available on their structure and composition. Though this can vary widely, we choose the well-documented reefs in Kaneohe Bay, Hawai'i as a basis for this modelling, which are primarily comprised of \emph{Porites compressa} coral \cite{KoehlHadfield2004}.
	
	On the assumption that the water's kinematic viscosity is $\nu \approx 10^{-6}\mathrm{m^2s^{-1}}$ and the flow within the reef has characteristic velocity scale $U = 10^{-2}\mathrm{ms^{-1}}$ \cite{KoehlHadfield2004}, the Reynolds number of the flow in the porous layer is
	\begin{equation}
		Re = \frac{Ul}{\nu} \approx 10^{4} l,
	\end{equation}
	where $l$ is a characteristic length scale for the problem -- the \emph{grain diameter} of the reef. Provided this diameter is of the order $10^{-3}\mathrm{m}$ (i.e. millimetre-scale), we can expect Darcy's law to be a valid description of the flow therein (Bear \cite{Bear1972} states that Darcy's law is a valid model for $Re \lesssim 10$). The reefs on Kaneohe Bay are especially simple to model, given that Koehl \& Hadfield \cite{KoehlHadfield2004} remark that waves break on the outside of the reef and then propagate steadily, and uniformly, over the water surface, satisfying our modelling assumptions. For the sake of simplicity, we start by modelling reefs where the permeability is a constant $\Pi\left(z\right) = \Pi_0$. The parameter values which are used in this model are summarised in table \ref{tab:params}.
	
	\begin{table}[H]
		\centering
		\begin{tabular}{cc}
			\hline
			\textbf{Parameter} & \textbf{Value} \\
			\hline
			Density of water, $\rho$ & $10^3\mathrm{kgm^{-3}}$ \\
			Dynamic viscosity of water, $\mu$ & $10^{-3}\mathrm{kg m^{-1}s^{-1}}$ \\
			Depth of water above reef, $d$ & $0.75\mathrm{m}$ \\
			Depth of reef, $D-d$ & $1.25\,\mathrm{m}$ \\
			Permeability of coral reef layer, $\Pi_0$ & $5\times10^{-7}\mathrm{m^2}$ \\
			Wave amplitude, $A$ & $0.05\mathrm{m}$ \\
			Wave frequency, $\omega$ & $2\mathrm{s^{-1}}$ \\
			\hline
		\end{tabular}
		\caption{Representative values of different parameters for the reef and the water flowing through it, determined through Koehl \& Hadfield \cite{KoehlHadfield2004} and private communication with Koehl (2019). Although it is difficult to determine the value of $D$, measurements suggest that the flow of water is negligible more than $2$ metres below the ocean surface.}
		\label{tab:params}
	\end{table}

	\subsection{Flow velocities}
	In this particular scenario, the numerically-calculated wavenumber is $k = 0.56 + 0.28\mathrm{i}$. As an initial check of the validity of our approach, we can compare the velocity magnitudes both above and within the reef as calculated by our model [using equations (\ref{eqn:upper_potential}) and (\ref{eqn:porous_velocities})] to the peak values presented in table 1 of \cite{KoehlHadfield2004}. This comparison is shown in table \ref{tab:compared_peaks}.
	\begin{table}[h]
		\centering
		\begin{tabular}{p{3cm}p{2cm}p{2cm}}
			\hline
			\textbf{Depth} & \textbf{Measured peak ($\mathrm{ms^{-1}}$)} & \textbf{Predicted peak ($\mathrm{ms^{-1}}$)} \\
			\hline
			$z=-0.48\mathrm{m}$ (above) & $0.12$ & $0.11$ \\
			$z=-0.91\mathrm{m}$ (within) & $0.03$ & $0.08$ \\
			\hline
		\end{tabular}
		\caption{A comparison of the peak horizontal velocities predicted by our modelling and reported in \cite{KoehlHadfield2004}, showing approximate agreement both above and within the reef layer.}
		\label{tab:compared_peaks}
	\end{table}

	\subsection{Stokes drift velocities}
\begin{figure}
		\centering
		\begin{subfigure}{\linewidth}
			\centering
			\includegraphics[width=\linewidth]{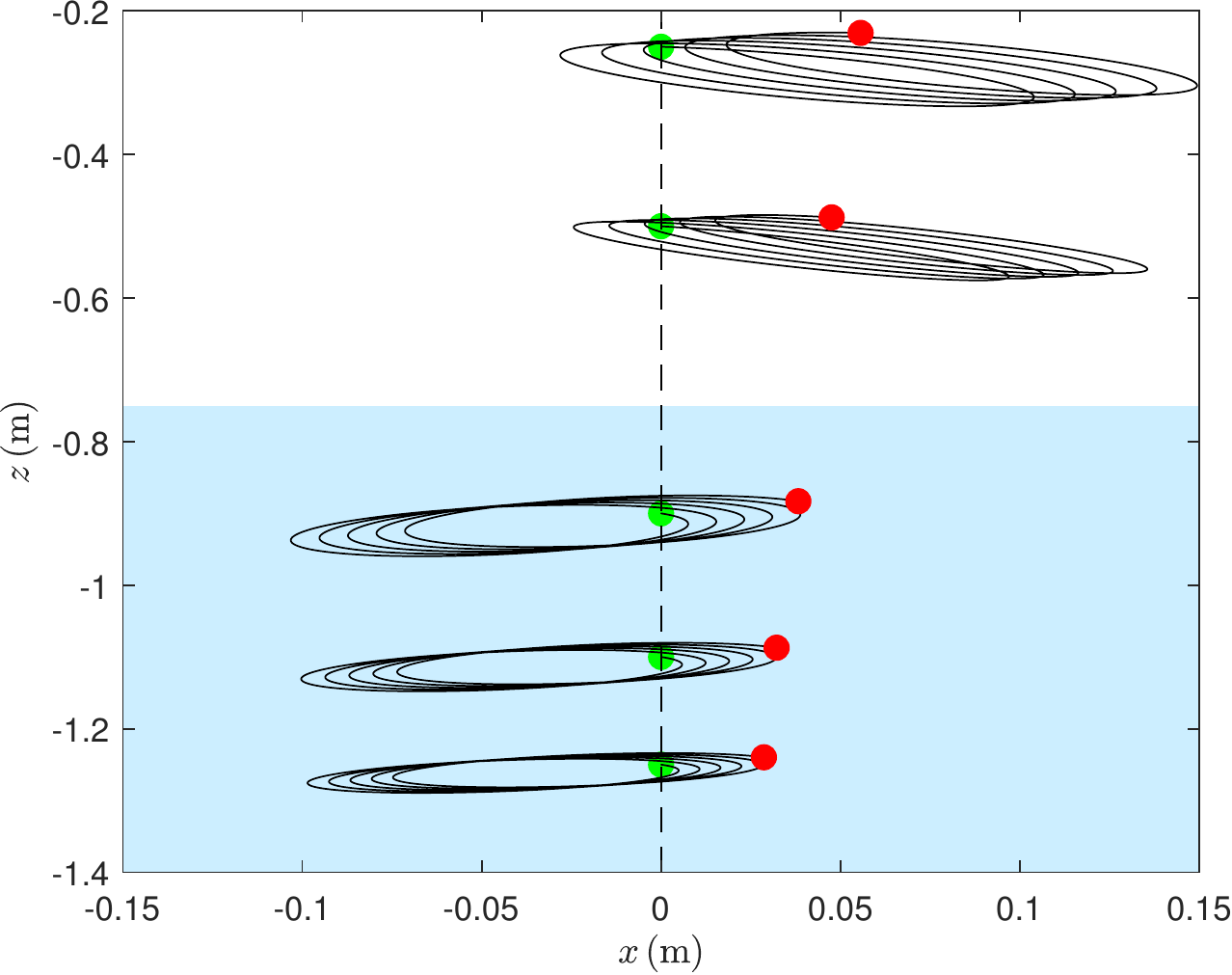}
		\end{subfigure}
		\par\bigskip
		\begin{subfigure}{\linewidth}
			\centering
			\includegraphics[width=\linewidth]{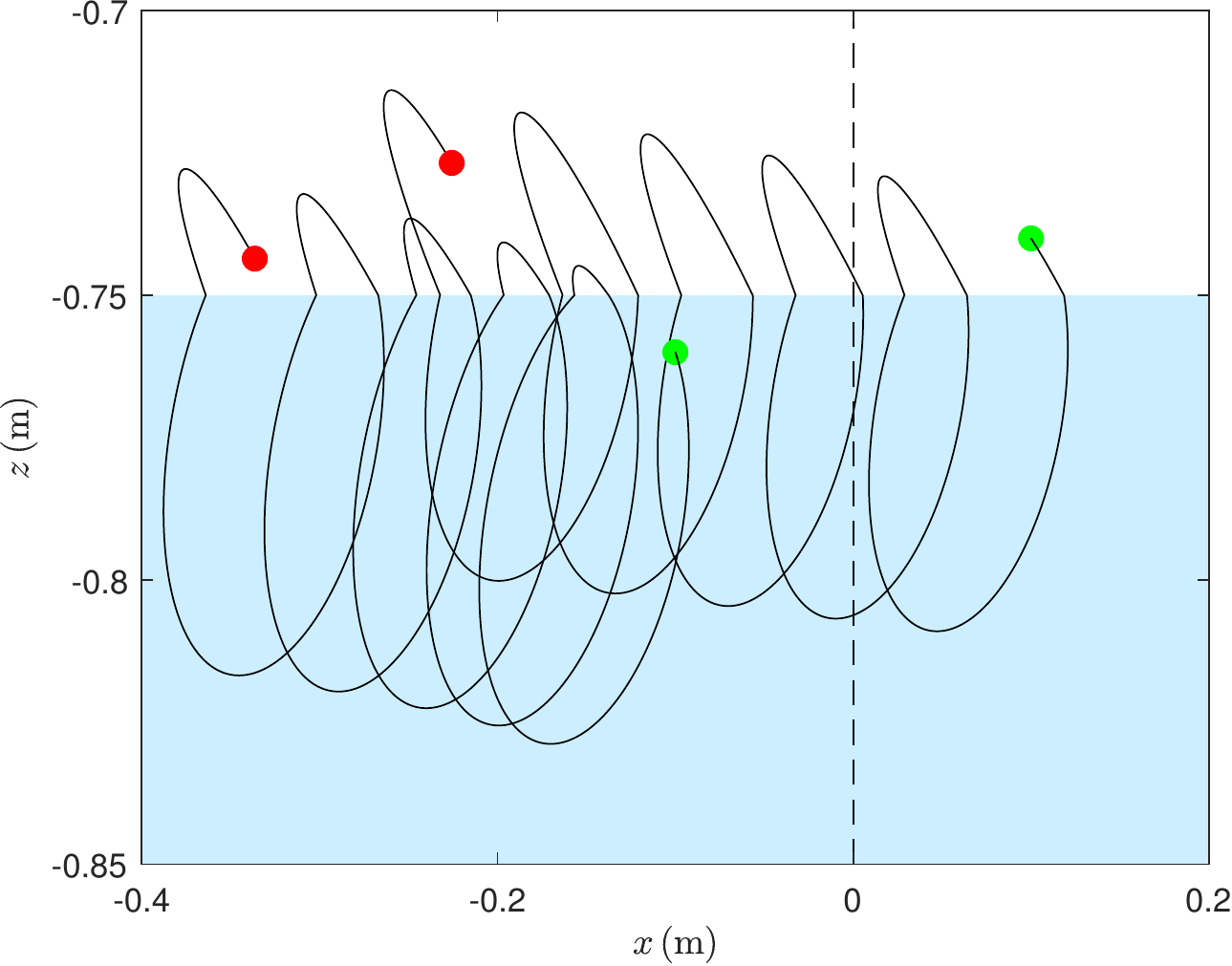}
		\end{subfigure}
		\caption{Paths of individual fluid `parcels' over 5 wave periods (i.e. from $t=0$ to $t=5\pi$), showing the net drift in their position. Starting positions are marked in green and final positions in red -- note the complicated paths taken by fluid parcels near the interface when they cross between regions.}
		\label{fig:tracePaths}
	\end{figure}

	Plotting the paths of fluid particles undergoing oscillatory motion in this model, as shown in figure \ref{fig:tracePaths}, we can see not only a drift effect in the positive $x$-direction (i.e. in the direction of wave propagation), as would be expected from Stokes' theory, but also a new vertical drift effect. This drift can be seen to arise from the damping of the waves, as the amplitude of vertical motion decreases with horizontal distance.
	
	As an individual fluid `parcel' moves forwards, it also moves downwards, before the direction of its horizontal velocity changes and it moves backwards, behind its original position. At this point, the parcel is moving upwards, but the magnitude of this upwards velocity is greater than that of the downward section of motion, owing to the damped amplitude. Therefore, over an entire period of motion, the fluid parcel not only experiences a net drift in the horizontal direction, but also a net upwards vertical drift.
	
\begin{figure}[htb]
		\centering
		\begin{subfigure}{\linewidth}
			\centering
			\includegraphics[width=\linewidth]{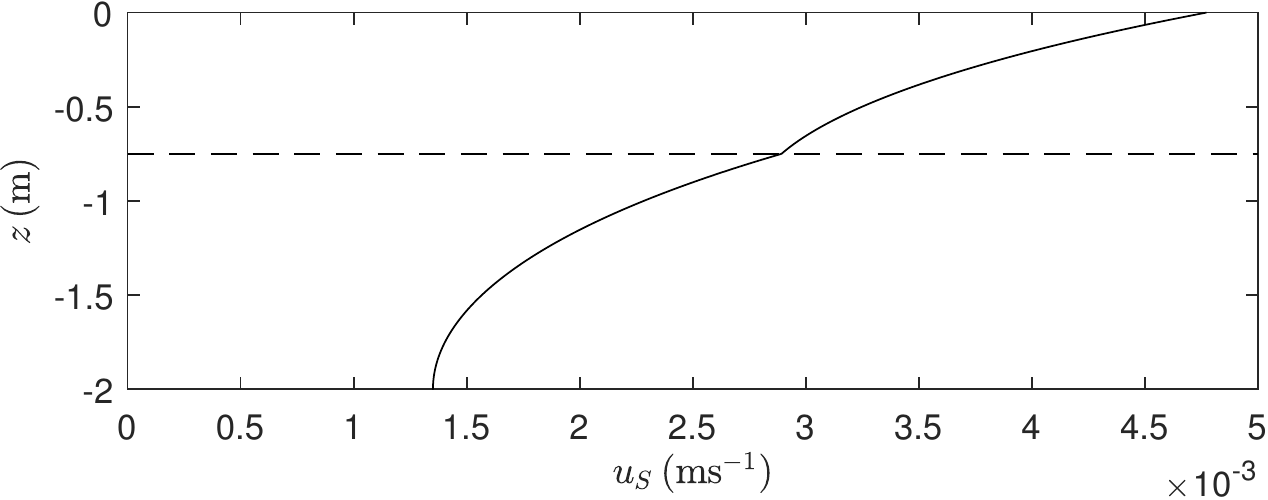}
		\end{subfigure}
		\par\bigskip
		\begin{subfigure}{\linewidth}
			\centering
			\includegraphics[width=\linewidth]{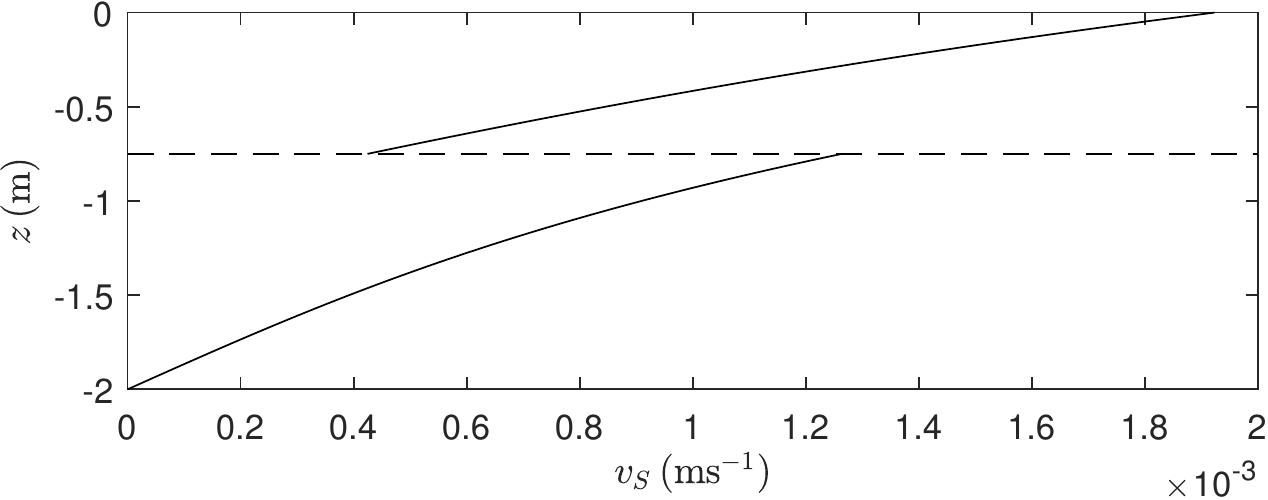}
		\end{subfigure}
		\caption{Plots of the horizontal and vertical Stokes drift velocities in the model with parameter values shown in table \ref{tab:params}.}
		\label{fig:profiles}
	\end{figure}

	Such a simplified model, however, breaks down close to the boundary between the reef and the fluid overlying it. An example of this behaviour is shown in the second plot of figure \ref{fig:tracePaths}, indicating that the expressions of equations (\ref{eqn:sd_upper}) and (\ref{eqn:sd_lower}) are no longer valid, and a direct numerical approach is required. In the case where these equations are valid, however, figure \ref{fig:profiles} shows representative horizontal and vertical drift velocities, of the order of a few millimetres per second, largely in agreement with the mean (drifting) velocities measured by Koehl \& Hadfield \cite{KoehlHadfield2004}.

	\subsection{Varying reef depths}
	It is apparent that changing the depth of the porous layer underlying the water will affect the amount by which waves are damped. Considering the model of table \ref{tab:params}, but varying the parameter $d$, it can be seen that the imaginary part of the wavenumber varies too, as shown in table \ref{tab:k}.
	
\begin{figure}[htb]
		\centering
		\begin{subfigure}{\linewidth}
			\centering
			\includegraphics[width=\linewidth]{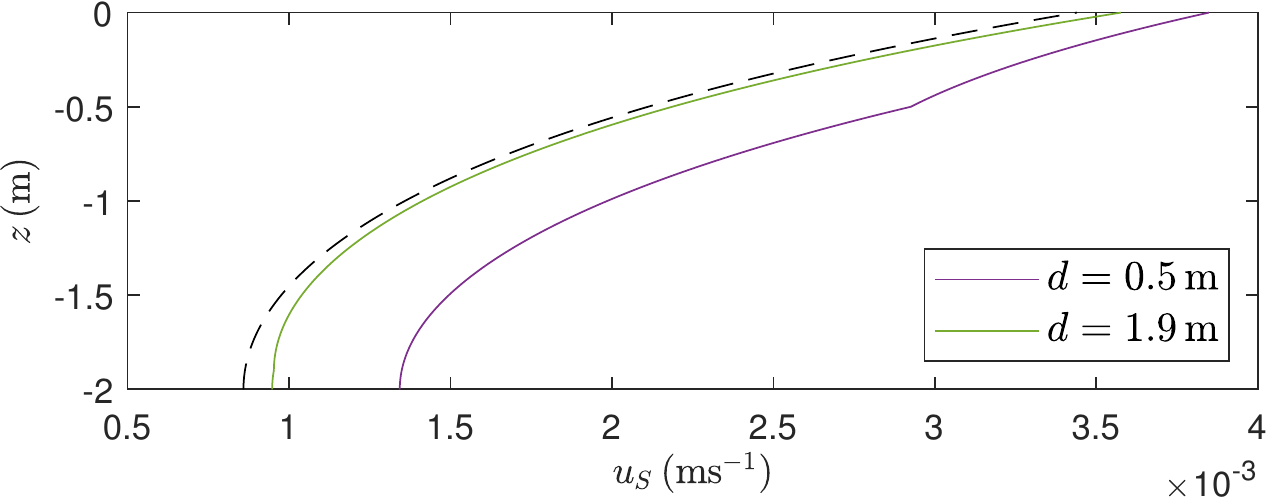}
		\end{subfigure}
		\par\bigskip
		\begin{subfigure}{\linewidth}
			\centering
			\includegraphics[width=\linewidth]{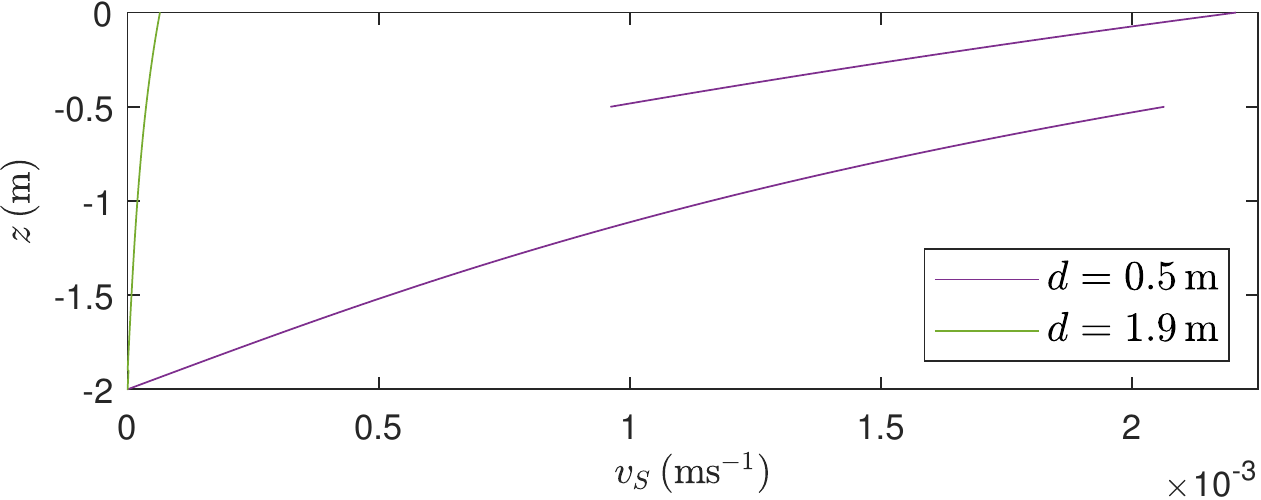}
		\end{subfigure}
		\caption{Plots of the horizontal and vertical Stokes drift velocities in the model with parameter values shown in table \ref{tab:params}, but allowing for the parameter $d$ to vary. The case of no porous layer is shown as a dashed line (there is no vertical drift with no porous layer).}
		\label{fig:comp_profiles}
	\end{figure}

	\begin{table}[H]
		\centering
		\begin{tabular}{ccc}
			\hline
			$\boldsymbol{d}\,\left(\mathrm{m}\right)$ & $\boldsymbol{k}\,\left(\mathrm{m^{-1}}\right)$ & \\
			\hline
			$0.10$ & $0.29 + 0.37\mathrm{i}$ & \\
			$0.25$ & $0.35 + 0.39\mathrm{i}$ & \\
			$0.50$ & $0.47 + 0.37\mathrm{i}$ & \\
			$1.00$ & $0.58 + 0.18\mathrm{i}$ & \\
			$1.50$ & $0.56 + 0.07\mathrm{i}$ & \\
			$2.00$ (no porous layer) & $0.52$ & \\
			\hline
		\end{tabular}
		\caption{Wavenumbers $k$ for different depths of coral reef in a shallow sea of total depth $D=2\,\mathrm{m}$, with all other parameters as in table \ref{tab:params}.}
		\label{tab:k}
	\end{table}
	This results in different drift velocities - two cases where $d=0.5\,\mathrm{m}$ and $d=1.9\,\mathrm{m}$ with all other parameters the same are shown in figure \ref{fig:comp_profiles}.
	
	\subsection{Role of spatial variation in permeability}
	Webber \& Huppert \cite{WebberHuppert2020} discuss in more detail the effect of taking the permeability to be a function of depth $\Pi\left(z\right)$ as opposed to a constant $\Pi_0$. Such a consideration is important physically, with marine reefs decreasing in permeability with depth. The dispersion relation is local, however, only dependent on the value of the permeability at the interface, and therefore any non-uniform permeability structure only manifests itself in the flows within the reef.
	
	The expressions for the Stokes drift velocity within the porous layer, equations (\ref{eqn:sd_lower}), show that permeability variations add an additional term in the horizontal drift, proportional to $\Pi'\left(z\right)$, when compared to expressions with constant $\Pi$. Assuming that the characteristic lengthscale of permeability variations is $L$,
	\begin{align}
		\frac{\frac{1}{2}\Pi' \sinh{\left(2\mathrm{Re}\left[k\left(z+D\right)\right]\right)}}{\Pi\, \mathrm{Re}\left[k\right]\cosh{\left(2\mathrm{Re}\left[k\left(z+D\right)\right]\right)}} \sim \frac{ \tanh{\left(2\mathrm{Re}\left[k\left(z+D\right)\right]\right)}}{L\mathrm{Re}\left[k\right]}
	\end{align}
	and so we can only reasonably neglect this term in cases where the permeability changes are slight or the waves are propagating at especially large or small wavenumbers.
	
	A second application of such spatial variation is seen when an algal turf layer overlies a coral reef -- this can have dramatic effects on flow within the reef itself and the habitat therein (Koehl \emph{et al.} \cite{KoehlPowellDobbins1997}). In this case, a thin layer of algal material, typically of lower permeability, overlies a reef which we assume to have uniform permeability. Restricting our attention to coralline algae, where issues of compression of the layer can be neglected \cite{RothEtAl2018}, we see that varying the permeability of this layer can have a dramatic effect on the transport, both horizontally and vertically, within the reef, as discussed in \cite{WebberHuppert2020}.
	
	\section{Conclusion}
	It has been shown that an analysis of surface gravity waves can be extended to a system where the fluid sits atop a saturated porous bed, incorporating the damping effect of this bed on the fluid motion. By matching a viscously-dominated Darcy flow law in this lower porous layer with inviscid potential flow above, a complete description of flow both above and within the bed can be derived, from which one can derive Stokes drift velocities in both layers.
	
	In addition to the well-known horizontal drift in the direction of wave propagation first derived in \cite{Stokes1847}, it is seen that the damping of the waves on the surface leads to a vertical drift in both layers. Though this model does not apply for fluid that crosses the boundary between the porous medium and the overlying water, in both regions the calculated drifts match measurements made by Koehl \& Hadfield \cite{KoehlHadfield2004} to a good degree of accuracy. Further field measurements in different reef locations around the world are needed to both confirm the generality of this result and provide more accurate parameter values for modelling.
	
	This different, and possibly simplified, model can be extended to consider permeable layers where the permeability varies with depth, as explored in more detail in Webber \& Huppert \cite{WebberHuppert2020}, and could then be further extended to consider other features of `real-world' coral reefs, including variable topographies and sloping sea floors, as well as incorporating existing work on the breaking of waves on reef lagoons and its effects on mass transport \cite{Monismith2007}. However, we would expect to still see this novel vertical drift effect as a major contributor to mass transport in reefs.
	
	\acknowledgements{Joseph Webber is thankful to the Heilbronn Fund at Trinity College, Cambridge, for funding much of his research into this topic. Both authors also thank Professor Mimi Koehl for a stimulating seminar in Cambridge that raised the problem considered in this paper, her continued encouragement and her provision of field measurements.}
	
	\appendix*
	\section{Appendix: Large-$\omega$ limit of $\mathrm{Im}\left(k\right)$}
	It is seen from (\ref{eqn:high_freq_lim}) that, to leading order, $k$ is real with value $\omega^2/g$ as $\omega \to \infty$, but it is also desirable to have a leading order approximation to $\mathrm{Im}\left(k\right)$ in this limit to understand the damping of the waves. Using the fact that $\tanh{x} \approx 1 - 2e^{-2x}$ as $x \to \infty$, (\ref{eqn:addition_formula}) becomes
	\begin{equation}
		J\left(1-2E_2\right) \approx \mathrm{i}\frac{\omega^2/gk - 1 + 2E_1}{1-\omega^2/gk + 2\omega^2 E_1 / gk},
	\end{equation}
	where $E_1 = e^{-2kd} \approx e^{-2\omega^2 d / g}$ and $E_2 = e^{-2k\left(D-d\right)} \approx e^{-2\omega^2 \left(D-d\right) / g}$ to leading order. We then postulate that
	\begin{equation}
		k = \frac{\omega^2}{g}\left(1 + K\right)\quad \text{with $K \ll 1$},
	\end{equation}
	and work to first order in $K$. This results in
	\begin{equation}
		J\left(1-2E_2\right)\left[K + 2\left(1-K\right)E_2\right] = \mathrm{i}\left(2E_1 - K\right).
	\end{equation}
	Solving for $\mathrm{Im}\left(K\right)$ gives
	\begin{equation}
		\mathrm{Im}\left(K\right) = \frac{2J\left(1-2E_2\right)}{1 + J^2 \left(1-2E_2\right)^4}\left(E_1 - 2E_1 E_2 + E_2\right).
	\end{equation}
	Finally, using the fact that $E_1 E_2 \ll E_1, E_2$ and $E_2 \ll 1$, it is found that
	\begin{equation}
		\mathrm{Im}\left(k\right) \approx \frac{2 J \omega^2}{g\left(1+J^2\right)}\left(E_1 + E_2\right)
	\end{equation}
	as $\omega \to \infty$, shown to be a good fit for even relatively small $\omega$ in figure \ref{fig:dispersion_plots}.
	
	\FloatBarrier
	\bibliographystyle{apsrev4-2}
	\bibliography{biblio}

\end{document}